\documentclass[pra,amssymb,aps,showpacs,amsmath]{revtex4}

\usepackage{graphicx}

\usepackage{latexsym}

\newcommand{\be}{\begin{equation}}
\newcommand{\ee}{\end{equation}}

\newcommand{\bex}{\begin{eqnarray}}
\newcommand{\eex}{\end{eqnarray}}
\newcommand{\bmin}{\begin{center}\begin{minipage}{460pt}}
\newcommand{\emin}{\end{minipage}\end{center}}
\begin{document}

\title{Unconditionally Secure Multipartite Quantum Key Distribution}
\author{Sudhir Kumar Singh}
\email{sudhirks@maths.iitkgp.ernet.in}
\affiliation{Dept. of Mathematics, IIT Kharagpur, Kharagpur- 721302, India.}
\author{R. Srikanth}
\email{srik@rri.res.in}
\affiliation{Raman Research Institute, Bangalore- 560080, Karnataka, India.}


\begin{abstract}
We consider the problem of secure key distribution among
$n$ trustful agents: the goal is to distribute an identical random bit-string
among the $n$ agents over a noisy channel
such that eavesdroppers learn little about it. We study
the general situation where the only resources required are
secure bipartite key distribution and authenticated classical communication.
Accordingly, multipartite quantum key distribution can be proven unconditionally
secure by reducing the problem to the biparitite case
and invoking the proof of security of bipartite quantum key distribution.
\end{abstract}

\maketitle

\section{Introduction}
With the growing use of the internet and other forms of electronic communication,
the question of secure communication becomes one of considerable importance.
Modern cryptographic techniques, based on the availability of ever increasing computational
power, and the invention of public key cryptography, provide practical solutions
for information security in various situations. But invariably these techniques 
are only computationally-- and not unconditionally-- secure, that is, they depend
on the (unproven) hardness of certain mathematical problems. As a result, it cannot
be guaranteed that future advances in computational power will not nullify their
cryptographic protection. Nevertheless, there does
exist a form of encryption with unconditional security: the use of one-time
key pads. These are strings of random numbers added by the information sender to
encode the message, to be subtracted by the receiver to decode.
Provided that the key material is truely random
and used only once, this system is unbreakable 
in the sense described by Shannon in the 1940s \cite{sha45}. It is critically
important that the pad is only used once, i.e., an encryption key can never be used 
twice. This restriction translates into the practical one of
key distribution (KD). This need to securely distribute the key between the users
makes it impractical in
many applications. Where it is used in real life (eg., in confidential communications
between governments), the one-time key pads are actually delivered in person by
some {\em trusted third party} \cite{sim00}, an arrangement prohibitively expensive 
for common usage and moreover not truely secure. Fortunately, recent
advances in quantum information theory have shown that unconditionally secure
key distribution is possible in principle.

That quantum information can, on account of quantum uncertainty and the no-cloning
principle \cite{woo82}, be used to distribute cryptographic
keys was realized two decades ago \cite{bb84,eke91}. More rigorous and
comprehensive proofs of this task, generally called quantum key distribution
(QKD), taking into consideration source, device and channel noise
as well as an arbitrarily powerful eavesdropper,
have been studied by various authors
\cite{may,bih,lochau,ina,sp00}. Recently, the issues of efficiency \cite{hwa03},
security in the presence of an uncharacterized source \cite{koa03} and
high bit-error rate tolerance \cite{wang} of QKD have been considered.
In particular, Lo and Chau \cite{lochau} showed that, given fault-tolerant
quantum computers, quantum key distribution over an arbitrarily long distance
of a realistically noisy channel can be made unconditionally secure.
This is a heartening development, since QKD is, among quantum
information applications, relatively easy to implement, and some large scale
implementations have already been achieved \cite{gis02,zei03}. The above mentioned
works consider QKD between two parties (ie., 2-QKD). It is of interest to
consider its extension to more than two parties (ie., $n$-QKD). 

The problem of $n$-QKD is to determine how $n$ parties, who are able to
communicate quantally, may share an identical and unconditionally secure, 
secret key among themselves in the presence of eavesdroppers. 
(A different generalization of 2-QKD gives multipartite quantum secret-sharing 
\cite{sca01}, which we do not consider here). 
In this work, we propose a protocol for this purpose and prove its unconditional security.
We note that a simple extrapolation of 2-QKD to $n$-QKD would suggest that the agents
should begin by sharing an $n$-partite entangled state. However, this proposition
suffers from two drawbacks: from a practical viewpoint, preparing $n$-partite
entanglement is no easy task; from a theoretical viewpoint, proving the
security of secure extraction of $n$ separated copies of a bit-string 
may not be simple, even given the existing proof of security of the bipartite case.
Our main result is that it is sufficient if some pairs of agents share
{\em bipartite} entanglement along any spanning tree connecting the $n$ agents,
who are taken to be vertices on a graph. 
In this way, $n$-QKD is reduced to a 2-QKD problem. Existing 2-QKD protocols 
\cite{may,bih,lochau,ina,sp00,hwa03,koa03,wang}
can be invoked to prove the unconditional security
of sharing nearly perfect Einstein-Podolsky-Rosen (EPR) pairs between two parties. 
They prescribe procedures for reliably sharing EPR pairs, and thence sharing
randomness, by virtue of fault-tolerant quantum computers, quantum error
correction and suitable random sampling. In the interests of brevity, we will
not elaborate these protocols here, and only point them out as subroutines for
the general $n$-QKD task.
Our protocol is fairly simple in the sense that the proof of its security
is built on top of the already proven security of the bipartite case. 
However, it is important to know the necessary and sufficient conditions 
on the network topology for our proposed protocol to work.

\section{Classical reduction of $n$-KD to 2-KD}\label{reduc}
As in 2-KD, the goal of $n$-KD is to show that 
$n$ trustful parties can securely share random, secret classical bits, even
in the presence of noise and eavesdropping. 
It is assumed that the $n$ agents can share authenticated
classical communication. It is convenient to treat the problem graph theoretically \cite{suds}.
The $n$ agents $A_i$ $(1 \le i \le n)$ are considered as the vertices (or nodes) 
of an undirected graph. An instance of a secure bi-partite channel being shared
between two parties is considered as an undirected edge between the two corresponding
vertices. A graph so formed is called a security graph.
It is obvious that if the security graph has a star topology (a hub vertex with
all edges radiating from it to the other vertices), a simple $n$-KD protocol
can be established. The agent at the hub vertex (say, called, Lucy) generates
a random bit string and transmits it to every other agent along the edges
to each of them. This will create a secure, identical random bit
string with each agent.

In real life situations, because of practical and geographical constraints, 
the $n$ agents may not form a security graph with
star topology. We describe a simple protocol that allows for more general
secure bi-partite connectivity between the agents. 
In particular, from among the secure bipartite channels suppose a spanning
tree (a graph connecting all vertices without forming a loop) 
can be constructed. This construction can be formalized in order to determine an 
optimal spanning tree. Some useful definitions are given below.

{\em Weighted security graph}:
Given $n$ parties treated as nodes on a graph, 
we extend the definition of a security graph 
to the {\em weighted security graph}.
A weight is associated with every edge and is defined 
to be some suitable measure of the cost of communicating by means of the
channel corresponding to the edge. 

{\em Minimum spanning security tree}: 
Consider the weighted security graph $G=(V,E)$. A spanning
tree selected from $G$, given by $G_1=(V,E_1), E_1 \subseteq E$ is called the
minimum spanning security tree if it minimizes the total weight of the graph. It need not be
unique and can be obtained 
using Kruskal's or Prim's algorithm \cite{cormen}. The minimum spanning security tree
minimizes the resources needed in the protocol as well as the
size of the sector eavesdroppers can potentially control.

{\em Terminal agent}: An agent that corresponds to a vertex of degree one 
(ie., with exactly one edge linked to it). On the other hand,
an agent that corresponds to a vertex of
degree greater than one is called a non-terminal agent.

We now present a classical subroutine that allows $n-1$ pair-wise shared random bits
to be turned into a single random bit shared between the $n$ parties.
\begin{enumerate}
\item 2-KD: Along the $n-1$ edges of a minimum spanning security tree, $n-1$ 
random bits are securely shared by means of some secure 2-KD protocol.
\item Each non-terminal agent $A_i$ announces his 
{\em unformly randomized record}:
this is the list of edges emanating from the vertex along with the corresponding
random bit values, to all of which 
a fixed random bit $x(i)$ is added.
\item This information is sufficient to allow every player, in conjunction with
her/his own random bit record, to reconstruct the random bits of all parties.
The protocol leader (say, Lucy) decides randomly on the terminal agent whose
random bit will serve as the secret bit shared among the $n$ agents.
\end{enumerate}
This subroutine consumes $n-1$ pair-wise shared random numbers to
give a one-bit secret key shared amongst the $n$-parties. To generate an
$m$-bit string shared among the $n$ agents, the subroutine is repeated $m$ times.

Given that the initial bipartite sharing of randomness is secure, we will
show that the above protocol subroutine allows some randomness to be shared between
the $n$ agents without revealing anything to an eavesdropper. It
involves each non-terminal agent announcing his uniformly randomized
record. Suppose one such, $A_i$, has the random record 0,1,1 on the three edges 
linked to his vertex. He may announce 0,1,1 (for $x(i)=0$) or 
1,0,0 (for $x(i)=1$). All the three agents linked
to him can determine which the correct string is by referring to their shared
secret bit. It is a straightforward exercise to see that each of other agents
linked to these three can determine the right bit string. Therefore, 
each agent can determine the random bits of all others.
Eavesdroppers, on the other hand, lacking knowledge of any
of the $n-1$ shared random bits, 
can only work out the {\em relative} outcomes of all parties. 
The result is exactly two possible configurations for each secret bit, 
which are complements of each other. The eavesdropper ``Eve" is thus maximally
uncertain about which the correct configuration is. 
Hence, Lucy's choice of a party to fix the secret bit reveals little to Eve.
Insofar as the $n$-parties are able to communicate
authenticated classical messages, the subroutine protocol is 
{\em as secure as} the underlying procedure for 2-KD.

It is obvious that the above protocol works for any spanning
security tree. Clearly, a sufficient condition for turning shared 
bipartite randomness into randomness shared between $n$ parties is that
the weighted security graph should contain at least one spanning
tree. On the other hand, if the security graph is disconnected, one easily
checks that it is impossible to arrive at a definite random bit securely
shared between both the disconnected pieces.
Therefore, the existence of at least one spanning tree in the weighted 
security graph is both a necessary and sufficient condition for the required task.

The amount of securely shared randomness may be quantified by the length
of shared
random bit string multiplied by the number of sharing agents. In the above protocol,
the $n-1$ instances of pair-wise shared randomness is consumed to produce
exactly one instance of a random bit shared between the $n$ parties. 
We can then define the `random efficiency' of the above protocol by 
$\eta = (n\times 1)/((n-1)\times2)$, 
which tends to (1/2) as $n \rightarrow \infty$.
Unconditional security of the above subroutine can in principle
only be guaranteed in a protocol which includes in step 1 a
quantum sub-routine that implements 2-QKD. In the following Section, we will
present one such, based on the Shor-Preskill protocols \cite{sp00}, as
an example.

\section{Quantum Protocol}

As in 2-QKD, the goal of the proposed $n$-QKD protocol is to show that 
$n$ trustful parties can securely distil random, shared, secret classical bits, whose
security is to be proven
inspite of source, device and channel noise and of Eve, an eavesdropper
assumed to be as powerful as possible, and in particular, having control over all
communication channels. From the result of the preceding Section, it follows that 
a quantum protocol is needed only in step 1 above. It will involve
establishing 2-QKD along a minimum spanning tree in order to 
securely share pair-wise randomness along spanning 
tree's edges and thence proceed to $n$-QKD. We assume as given
the security of establishing pair-wise randomness along a spanning tree by means of 
a quantum communication network, based on a secure 2-QKD protocol
\cite{may,bih,lochau,sp00,ina,hwa03,koa03,wang}. In principle, these protocols guarantee
security under various circumstances. 

In an $n$-QKD scheme, the insecurity of even one of the players
 can undermine all. Hence
additional classical processing like key reconciliation and privacy
amplification \cite{cis76} of the final key may be needed at the $n$-partite level. 
In the full $n$-QKD protocol that we present below,
following Ref. \cite{sp00} we exploit the connection of
error correction codes \cite{mac} with key reconciliation and privacy amplification.
These procedures have been 
extensively studied by classical cryptographers \cite{gis02}, and other
possibilities exist.

In particular, we adopt a quantum protocol wherein pair-wise randomness is
created by means of sharing EPR pairs (this follows the pattern set by the
Ekert \cite{eke91}, Lo-Chau \cite{lochau} and Modified Lo-Chau \cite{sp00} protocols, but 
entanglement is not necessary, as seen in the original BB84 protocol).
The basic graph theoretic definitions introduced above apply also for the quantum
case, except that now the security channels correspond to shared EPR pairs.
In place of a secure bipartite channel, an instance of EPR pair shared
between two parties is considered as an undirected edge between the two corresponding
vertices. A graph so formed is called an EPR graph \cite{suds}.
The analog of the weighted security graph is the
{\em weighted EPR graph}, and that of the minimum spanning security tree is the
{\em minimum spanning EPR tree}. 
Let us enumerate the $n$ parties as $A_1,A_2,\cdots,A_n$.
Suppose that only $A_{i_1},A_{i_2},\cdots,A_{i_s}$ $(i_1,i_2,\cdots,i_s\in
\{1,2,\cdots,n \})$ are capable of producing EPR pairs and
$S \equiv \{A_{i_1},\cdots,A_{i_s}\}$ is the set of all such vertices,
with $S\ne \emptyset$.
We construct a weighted undirected graph $G=(V,E)$ as one whose every edge
must contain a vertex drawn from the set $S$, as follows:
$V\equiv \{A_i;~i=1,2,\cdots,n\}$ and
$E \equiv \{(A_i,A_j) ~\forall~ A_i \in S ~{\rm and}~ \forall~ A_j \in V;~
i\neq j\}$. And the weight of edge $(A_i,A_j)$ is defined to be 
$w_{i,j} \propto$ number of quantum repeaters \cite{repeater}
(more generally: entanglement distilling resources \cite{cirnat}) 
required to be put between $A_i$ and $A_j$. 
Usually, the larger the distance between two
agents, the larger is the weight.
 The minimum spanning EPR tree minimizes the number 
of quantum repeators needed, and, in general, the resources needed in the protocol
(EPR pairs, etc.) subject to the constraint of available EPR sources. Apart from 
improving efficiency in terms of costs incurred,
this optimization is also important from the security perspective
in that it minimizes the size of the sector that Eve can potentially control.

Let ${\cal C}$ be a classical $t$-error correcting $[m,k]$-code \cite{mac}. 
We now present a protocol that consumes $n-1$ pair-wise securely shared 
sets of EPR pairs to create random bits shared between the $n$ parties with
asymptotic efficiency $\eta = (1/2)k/m$, where $k/m$ is the rate of the code.
The classical subroutine described in the previous Section is adapted to
include key reconciliation and privacy amplification at the $n$-partite level,
that uses the group theoretic properties of ${\cal C}$.
\begin{enumerate}
\item EPR protocol:
Along the $n-1$ edges of the minimum spanning EPR tree, EPR pairs
are shared (using eg., the Lo-Chau \cite{lochau}
or Modified Lo-Chau protocols \cite{sp00}). 
Let the final, minimum number of EPR pairs distilled along any edge of the 
minimum spanning EPR tree be $2m$. A projective measurement in the computational 
basis is performed  by all the parties on their respective qubits to obtain
secure pair-wise shared randomness along the tree edges
(making due adjustments according to whether the entangled spins are correlated or
anti-correlated).
\item Classical subroutine of Section \ref{reduc}:
All non-terminal vertices announce their 
{\em unformly randomized outcome record}.
This information in principle allows every party, in conjunction with
her/his outcome, to reconstruct the outcomes of all other parties, save for some errors
of mismatch. 
\item For each set of $n-1$ shared EPR pairs, 
protocol leader Lucy decides randomly on the terminal party whose outcome
will serve as the secret bit.
\item Lucy decides randomly a set of $m$ bits to be used as
check bits, and announces their positions.
\item All parties announce the value of their check bits. If too few of
these values agree, they abort the protocol.
\item Lucy broadcasts $c_i \oplus v$, where  
$v$ is the string consisting of  the remaining code (non-check) bits, and
$c_i$ is a random codeword in ${\cal C}$.
\item Each member $j$ from amongst the remaining $n-1$ parties 
subtracts $c_i \oplus v$ from his
respective code bits, $v\oplus\epsilon_j$, and corrects the result,
$c_i \oplus \epsilon_j$, to a codeword in ${\cal C}$. Here $\epsilon_j$
is a possibly non-vanishing error-vector.
\item The parties use $i$ as the key.
\end{enumerate}

A rigorous proof of the security of the $n$-QKD scheme
requires: (a) the explicit construction of a procedure such that 
whenever Eve's strategy has a non-negligible probability of
passing the verification test by the $n$ parties, 
her information on the final key will be exponentially small. 
(b) the shared, secret randomness is robust against source, device
and channel noise. By construction, our scheme combines a
2-QKD scheme to generate pair-wise shared randomness and a classical
scheme to turn this into multipartite-shared randomness. The security
of the latter (in its essential form)
was proven in Section \ref{reduc}. Therefore the security of
the protocol with respect to (a) and (b) reduces to that of the 
2-QKD in step 1.
For various situations, 2-QKD can be secured, as proven in Refs
\cite{may,bih,lochau,ina,sp00}. 
For example, Lo and Chau \cite{lochau} and Shor and Preskill \cite{sp00}
have proved that EPR pairs can be prepared to be nearly perfect, even in the
presence of Eve and channel noise. 
Their proofs essentially relies on the idea
that sampling the coherence of the qubits allows one to place an upper bound on the
effects due to noise and information leakage to Eve.
Yet, subject to the availability of high quality
quantum repeaters and fault-tolerant quantum computation, in principle 
2-QKD can be made unconditionally secure \cite{lochau}.

In regard to the key reconciliation part:
in step (3), each non-terminal vertex party announces his uniformly randomized
outcome record. Here this consumes $m$ instances of $n-1$ pair-wise shared random
bits into $k$ random bits shared between the
$n$ agents while revealing little to Eve. The random efficiency is given by
\mbox{$\eta = (k \times n \times 1)/(m \times (n-1) \times 2)$}, which tends to
$(1/2)k/m$ as $n \rightarrow \infty$.
The check bits, whose positions and values are announced in steps (4) and (5), 
are eventually discarded.
Steps (7) and (8) involve purely local, classical operations \cite{lokal}.
If security of step (1) against Eve is guaranteed, the string $v$,
and thereby the string $c_i \oplus v$ announced by Lucy in step (6), 
are completely random, as far as Eve
can say. So, she (Eve) gains nothing therefrom.
Hence her mutual information with any of the $n-1$ (sets of)
random bits does not increase beyond
what she has at the end of the EPR protocol \cite{clauthen}. 

Finally, step (5) permits with high probability to determine whether the key
can be reconciled amongst the $n$ players.
The check bits that the parties measure behave
like a classical random sample of bits \cite{sp00}. 
We can then use the measured error rates
in a classical probability estimate. For any two parties,
the probability of obtaining
more than $(\delta + \epsilon)n$ errors on the code bits and fewer than 
$\delta n$ errors on the check bits is asymptotically less than
$\exp[-0.25\epsilon^2n/(\delta-\delta^2)]$. Noting that the errors
on the $n$ check vectors 
are independent, it follows that probability that the check vectors
are all scattered within a ball of radius $\delta n$ but one or more code
vectors fall outside a scatter ball of radius $(\delta+\epsilon)n$ is 
exponentially
small, and can be made arbitrarily small by choosing sufficiently small $\delta$.
The decision criterion adopted in step (5) is calculated so that 
the Hamming weight of the error vectors $\epsilon_j$ estimated in the above fashion
will be less than $t$ with high probability. Hence all parties correct 
their results to the same codeword $c_i$ in step (8) with high probability.
This completes the proof of unconditionally security of $n$-QKD.


\acknowledgments
We thank Prof. Anil Kumar, Prof. J. Pasupathy and Ritesh Kumar Singh for discussions.
SKS thanks Prof. Anil Kumar for enabling his visit to IISc during which this
work was done. RS's work was partially supported by DRDO
project 510 01PS-00356.



\begin{thebibliography}{100}
\bibitem{lokal} Unless the privacy of local operations is guaranteed, 
the matter of security is quite hopeless.
\bibitem{clauthen} In particular, no quantum key distribution scheme can 
succeed if Eve has the power to impersonate agent $A_i$ while communicating to
$A_j$ and to impersonate $A_j$ while communicating to $A_i$. To address 
this problem, $ A_i$ and $A_j$ will need to authenticate their communications
(cf. Ref. \cite{lochau} for details).
\end{thebibliography}
\end{document}